\documentstyle[twocolumn,aps,prl,psfig]{revtex}
\begin{document}
\draft

\title{$A$-dependence of nuclear transparency in quasielastic
$A(e,e'p)$\\
at high $Q^2$}

\author{
T.~G.~O'Neill,$^1${$^,$}{\cite{atanl}}\,
W.~Lorenzon,$^1${$^,$}{\cite{atpenn}}\,
P.~Anthony,$^2$\,
R.~G.~Arnold,$^3$\,
J.~Arrington,$^1$\,
E.~J.~Beise,$^1${$^,$}{\cite{atmaryland}}\,
J.~E.~Belz,$^1${$^,$}{\cite{atboulder}}\,
P.~E.~Bosted,$^3$\,
H.-J.~Bulten,$^4$\,
M.~S.~Chapman,$^5$\,
K.~P.~Coulter,$^6${$^,$}{\cite{atmich}}\,
F.~Dietrich,$^2$\,
R.~Ent,$^5${$^,$}{\cite{atcebaf}}\,
M.~Epstein,$^7$\,
B.~W.~Filippone,$^1$\,
H.~Gao,$^1$\,
R.~A.~Gearhart,$^8$\,
D.~F.~Geesaman,$^6$\,
J.-O.{} Hansen,$^5$\,
R.~J.~Holt,$^6$\,
H.~E.~Jackson,$^6$\,
C.~E.~Jones,$^4${$^,$}{\cite{atanl}}\,
C.~E.~Keppel,$^3${$^,$}{\cite{atvirgin}}\,
E.~R.~Kinney,$^{9}$\,
S.~Kuhn,$^{10}${$^,$}{\cite{atodu}}\,
K.~Lee,$^5$\,
A.~Lung,$^3${$^,$}{\cite{atcaltech}}\,
N.~C.~R.~Makins,$^5$\,
D.~J.~Margaziotis,$^7$\,
R.~D.~McKeown,$^1$\,
R.~G.~Milner,$^5$\,
B.~Mueller,$^1$\,
J.~Napolitano,$^{11}$\,
J.~Nelson,$^5${$^,$}{\cite{atslac}}\,
V.~Papavassiliou,$^6$\,
G.~G.~Petratos,$^8${$^,$}{\cite{atkent}}\,
D.~H.~Potterveld,$^2$\,
S.~E.~Rock,$^3$\,
M.~Spengos,$^3$\,
Z.~M.~Szalata,$^3$\,
L.~H.~Tao,$^3$\,
K.~van~Bibber,$^2$\,
J.~F.~J.~van~den~Brand,$^4$\,
J.~L.~White,$^3$\,
B.~Zeidman$^6$}

\address{\hss\\
$^1$\ California Institute of Technology, Pasadena, California
91125\\
$^2$\ Lawrence Livermore National Laboratory, Livermore, California
94550\\
$^3$\ American University, Washington, D. C. 20016\\
$^4$\ University of Wisconsin, Madison, Wisconsin 53706\\
$^5$\ Massachussetts Institute of Technology, Cambridge,
Massachusetts 02139\\
$^6$\ Argonne National Laboratory, Argonne, Illinois 60439\\
$^7$\ California State University, Los Angeles, California 90032\\
$^8$\ Stanford Linear Accelerator Center, Stanford, California
94309\\
$^{9}$\ University of Colorado, Boulder, Colorado 80309\\
$^{10}$\ Stanford University, Stanford, California 94305\\
$^{11}$\ Rensselaer Polytechnic Institute, Troy, New York 12180\\
\hss\\}
\date{\today}
\maketitle

\begin{abstract}
The $A$-dependence of the quasielastic $A(e,e'p)$ reaction has been
studied at SLAC with ${}^2$H, C, Fe, and Au nuclei at momentum
transfers $Q^2 = 1$, 3, 5, and 6.8~(GeV/c)$^2$. We extract the
nuclear transparency $T(A,Q^2)$, a measure of the average probability
that the struck proton escapes from the nucleus $A$ without
interaction. Several calculations predict a significant increase in
$T$ with momentum transfer, a phenomenon known as Color Transparency.
No significant rise within errors is seen for any of the nuclei
studied.
\end{abstract}
\pacs{PACS numbers: 25.30}

\narrowtext
In 1982, Mueller and Brodsky \cite{mb82} proposed that in wide angle
exclusive processes, the soft initial and final state interactions
(ISI and FSI) of hadrons in nuclei would vanish at high energies.
This effect, originally based on arguments using perturbative QCD, is
called ``Color Transparency'' (CT), in reference to the disappearance
of the color forces between the hadrons and nuclei. Evidence for the
CT effect can be sought by measurement of the nuclear transparency
$T$, defined as the ratio of the measured cross section to the cross
section expected in the limit of complete CT (i.e., no ISI or FSI),
as a function of the 4-momentum transfer squared, $Q^2$, and nuclear
mass, $A$. For CT to be observable in quasielastic $A(e,e^\prime p)$
scattering, the recoiling proton must maintain its reduced
interaction with other nucleons over a distance comparable to the
nuclear radius. This is probed directly by measuring the $A$
dependence of $T$. At low energies, $T < 1$ because of absorption or
deflection of the hadrons by ISI and FSI with the nucleus. As the
energy increases, and if CT effects begin to dominate the scattering,
$T$ should increase towards unity \cite{farrar88}. Some recent models
of CT predict significant increases in $T$ for $Q^2$ as low as
5~(GeV/c)$^2$ \cite{farrar88,jenmil91,benhar92,fsz93,ralston88}. We
present measurements of $T$ for the reaction $A(e,e'p)$ on ${}^2$H,
C, Fe, and Au nuclei at $Q^2 = 1$, 3, 5, and 6.8~(GeV/c)$^2$.

The first experiment to investigate CT was performed by Carroll {\it
et al.} \cite{carrol88} using simultaneous measurements of $A(p,2p)$
and H$(p,2p)$ reaction rates at Brookhaven National Laboratory. Their
results showed $T$ increasing for $Q^2\simeq \hbox{3--8~(GeV/c)}^2$,
but then decreasing for $Q^2 \simeq \hbox{8--11~(GeV/c)}^2$. Because
of the subsequent decrease, the rise at lower momentum transfer
cannot be taken as an unambiguous signal of CT. Ralston and Pire
\cite{ralston88} suggest that the maximum in $T$ is due to a soft
process that interferes with the perturbative QCD amplitude in free
proton-proton scattering but is suppressed in the nuclear
environment. Such ambiguities should be smaller in $A(e,e'p)$
reactions because of the simplicity of the elementary electron-proton
interaction compared to the proton-proton interaction.

The experiment reported here was performed in End Station A at SLAC
using the electron beam from the Nuclear Physics Injector
\cite{npas84}. Details of the experiment have been published
previously \cite{makins94}. Kinematics for the present data are shown
in Table~I. Solid targets of 2\% (C), 6\% (C, Fe, and Au), and 12\%
(Au) radiation length and liquid targets of 4.0 (${}^1$H and ${}^2$H)
and 15.7~cm (${}^2$H) were used. The angle of the proton spectrometer
was varied to account for the Fermi motion of the initial proton
(so-called perpendicular kinematics).

Measurement of the electron and proton in coincidence allows
reconstruction of the ``missing'' energy, $E_m \equiv \nu - E_p' +
M_p - K_{A-1}$, and momentum, ${\bf p}_m \equiv {\bf p'} - {\bf q}$,
not accounted for in the detected particles \cite{fmoug83}. In the
Plane Wave Impulse Approximation (PWIA), these are equal to the
separation energy $E_s$ and momentum ${\bf p}$ of the struck proton,
which has initial 4-momentum $p \equiv (M_p - E_s - K_{A-1}, {\bf
p})$. Here $q = (\nu, {\bf q})$ is the virtual photon 4-momentum
transfer ($Q^2 \equiv -q^2$), $p' = (E_p', {\bf p'})$ is the
4-momentum of the detected proton, and $K_{A-1}$ is the kinetic
energy of the recoiling $A-1$ system.

We define the nuclear transparency $T$ as the ratio of the measured
coincidence rate to the rate calculated in the PWIA. The PWIA
quasielastic cross section is \cite{def83}
\begin{equation}
{d^6 \sigma \over
dE_e' d\Omega_{e'} dE_p' d\Omega_{p'}} =
p' E_p' \sigma_1^{cc} S({\bf p}, E_s)\;.
\label{eq:sig6}
\end{equation}
Here $dE_e' d\Omega_{e'}$ and $dE_p' d\Omega_{p'}$ refer to the
outgoing electron and proton, respectively. The nuclear structure is
characterized by the spectral function $S({\bf p}, E_s)$, the
probability density for finding a proton with separation energy $E_s$
and 3-momentum ${\bf p}$. The electromagnetic interaction is
specified by $\sigma_1^{cc}$ \cite{def83}, the square of the elastic
scattering amplitude of an electron and a moving off-shell proton.
Other forms for this amplitude, including the on-shell value, have
been tested with little ($\leq2\%$) effect on the measured $T$. The
dipole form for $G_E^p$ and the Gari-Kr\"{u}mpelmann \cite{gari85}
form for $G_M^p$ are assumed.

Details of the Monte Carlo program used to compute the PWIA
cross-section were presented in a previous publication
\cite{makins94}. In the present analysis, we use a delta function for
the ${}^1$H spectral function and determine the ${}^2$H spectral
function using the full Bonn potential \cite[Table~II]{bonnxx}. For
the solid targets, we use Independent Particle Shell Model (IPSM)
spectral functions; the energy levels are characterized by a
Lorentzian energy profile (due to the finite lifetime of the one-hole
state), and the momentum distributions are calculated using
Woods-Saxon nuclear potentials with shell-dependent parameters. The
Lorentzian and Woods-Saxon parameters are determined from fits to
spectral functions extracted from previous $A(e,e'p)$ experiments
(Ref.~\cite{fmoug83} for C and Fe, Ref.~\cite{Quint} for Au).
Descriptions of the deepest-lying shells of Fe and Au were taken from
a Hartree-Fock calculation \cite{Negele} since data on these shells
are inconclusive. For Fe and Au, the spectral function parameters
were varied to provide better agreement with the $Q^2 = 1$ and $Q^2 =
3~\hbox{(GeV/c)}^2$ data of the present experiment \cite{ton94}. The
uncertainty in the spectral function parameters results in 2\%
systematic uncertainties in $T$ for C, 3\% for Fe, and 5\% for Au.
The IPSM spectral function does not include the effects of
short-range nuclear correlations, which move strength to $p_m$
greater than the Fermi momentum. The measured $T$ must be corrected
by the ratio of $\int S\, d^3p\, dE_s$ for the correlated and the
IPSM spectral functions, integrated over the measured $E_m$ and $p_m$
range. For C, the correction factor is $1.11 \pm 0.03$, inferred from
${}^{12}$C \cite{sick} and ${}^{16}$O \cite{vanorden} spectral
functions that include the effects of correlations. For Fe and Au we
use a correlated nuclear matter spectral function corrected for
finite nucleus effects \cite{ji,liuti94}, yielding correction factors
of $1.22 \pm 0.06$ for Fe and $1.28 \pm 0.10$ for Au.

The data used to extract $T$ are restricted to a kinematic region
where the spectrometer acceptances and the shape of the spectral
function are well understood. The acceptance of each spectrometer is
restricted to $\pm 5\%$ of the central momentum, $\pm 15$~mr in
in-plane angle, and $\pm 40$~mr in out-of-plane angle. Furthermore,
we require $-30 < E_m < 100$~MeV (negative $E_m$ account for finite
resolution effects) and restrict the range of $p_m$. By eliminating
events with $E_m \gtrsim 140~\hbox{MeV} \simeq m_\pi$, we ensure that
no inelastic processes have occurred. For ${}^1$H and ${}^2$H, we use
$p_m<170~\hbox{MeV/c}$. For the C, Fe, and Au targets, we use a range
in $p_m$ that provides uniform coverage over all $Q^2$: $0 < p_m <
250~\hbox{MeV/c}$ \cite{pm} for Fe and C and $0 < p_m <
210~\hbox{MeV/c}$ for Au because fewer recoil proton angles were
measured for this target. The transparency at each $Q^2$ is the
weighted average of $T$ over the proton spectrometer angle settings.
The resulting $T$ is insensitive at the $\sim5\%$ level to variations
in the above kinematic limits.

\psfig{file=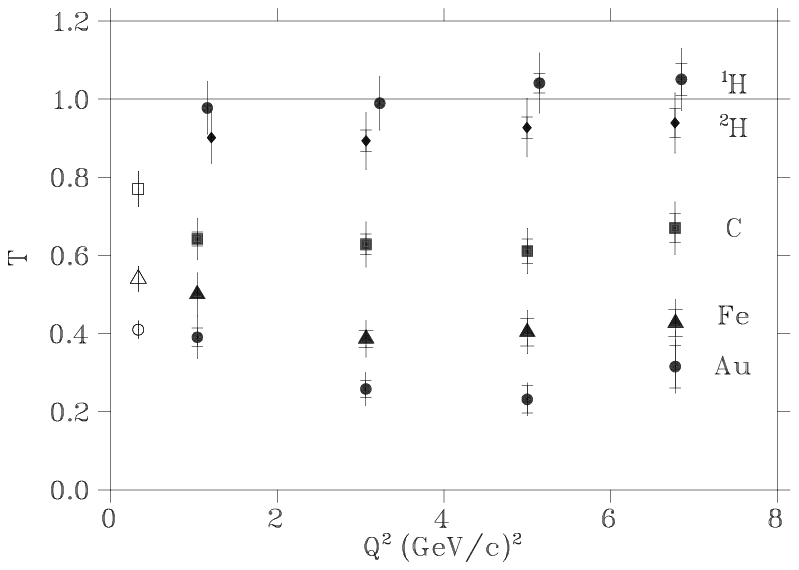}
\vbox{\footnotesize Figure 1. Nuclear transparency for $A(e,e'p)$ as
a function of $Q^2$. The inner
error bars are the statistical uncertainty, and the outer error bars
are the
statistical and systematic uncertainties added in quadrature. The
points at
$Q^2=0.33~\hbox{(GeV/c)}^2$ are from Ref.~\protect\cite{garino92} for
C, Ni and
Ta targets.}
\vspace{1cm}

Figure 1 shows the measured transparency as a function of $Q^2$. Note
that the results for ${}^{12}$C differ slightly (2--3\%) from those
previously published \cite{makins94}, principally due to improvements
in the radiative corrections \cite{ton94}. The ${}^1$H results are
consistent with the expected $T = 1$ (no absorption), while the
${}^2$H transparencies appear to be systematically below unity by
$\sim 8\%$. For the $A > 1$ targets at all $Q^2$, the measured $p_m$
and $E_m$ distributions are in reasonable agreement
\cite{makins94,ton94} with those calculated in the PWIA model. As
this comparison is made using a single spectral function for each
nucleus (renormalized at each $Q^2$ by the measured transparency
$T$), it indicates that the PWIA description of quasielastic
scattering is valid at $Q^2 \geq 1~\hbox{(GeV/c)}^2$.

Fractional systematic uncertainties in $T$ include 3\% for detection,
tracking, and coincidence timing; 5\% for spectrometer acceptances;
2\% for proton absorption; $\leq 0.9\%$ for charge, target
thicknesses, and dead time; 3\% for radiative effects; 2\% for
$G_E^p$ and $G_M^p$ parametrization; 2\% for $\sigma_1^{cc}$ (except
for ${}^1$H); 2--5\% for $S({\bf p}, E_s)$ (solid targets only); and
3--8\% for the correlation correction (solid targets only). Color
Transparency is expected to produce an increase in $T$ with
increasing $Q^2$ for the $A > 1$ targets. There is no evidence within
experimental errors of such an increase in the measured $Q^2$ range.
The rise in the value of $T$ at $Q^2 \leq 1\hbox{(GeV/c)}^2$
(including the data from Ref.~\cite{garino92}) is at least partially
due to the smaller nucleon-nucleon total cross section at momenta
$\simeq1\hbox{GeV/c}$, as has been suggested in Ref.~\cite{fsz93}.
For $Q^2 \geq 3~\hbox{(GeV/c)}^2$, the magnitude of the measured $T$
is within the range of the existing Glauber model calculations (i.e.,
no CT effects)
\cite{farrar88,jenmil91,benhar92,fsz93,seki93,rinat93,nik93}).

To combine the results from different nuclei and improve the
sensitivity to CT effects, we can use a simple model for the
$A$-dependence (for $A \geq 12$) of the transparency to obtain an
effective nucleon-nucleon cross section ($\sigma_{\rm eff}$) for each
momentum transfer. This model assumes classical attenuation for the
proton propagating in the nucleus with a $\sigma_{\rm eff}$ that is
independent of density:
$$ T_{\rm class} = {1 \over Z}\int d^3r \, \rho_Z
({\bf r})\exp{ [-\int dz' \, \sigma_{\rm eff} \rho_{A-1} ({\bf
r'})]}\;.
$$

In the limit of complete CT, one would expect $\sigma_{\rm eff}
\rightarrow 0$. For this calculation, the nuclear density
distributions were taken from Ref.~\cite{deVries} and $\sigma_{\rm
eff}$ is the only free parameter. We also assume that the hard
scattering rate is accurately described at each $Q^2$ by our PWIA
model, unlike Ref.~\cite{ralston93}, where the hard scattering
amplitude was also varied as a free parameter. The results of fitting
this model to the measured transparency for the C, Fe, and Au targets
is shown in Fig.~2 (solid curve). Also shown (dashed curve) is a
simple $T = A^{\alpha}$ parameterization, where complete CT would
correspond to $\alpha = 0$. The classical attenuation model provides
a reasonable parameterization of the data (somewhat better than the
$A^{\alpha}$ fits) and the fitted values of $\sigma_{\rm eff}$ are
tabulated in Table~II, where one observes a decrease in $\sigma_{\rm
eff}$ at $Q^2=1~\hbox{(GeV/c)}^2$ correlated with the measured
decrease in the free nucleon-nucleon cross section. We note that
$\sigma_{\rm eff}$ is noticeably lower than the free cross section
$\sigma_{\rm free}$ (Table~II), as could be expected from quantum
effects not accounted for in the classical calculation, as well as
nuclear effects such as Pauli blocking, short-range correlations,
etc. \cite{vj92}, which are important effects at lower $Q^2$. In
addition, the finite experimental acceptance has been shown
\cite{fsz93,nik93} to account for some of this effect. The ratio of
$\sigma_{\rm eff}$ to $\sigma_{\rm free}$ is consistent with a
constant value of 0.68.

\psfig{file=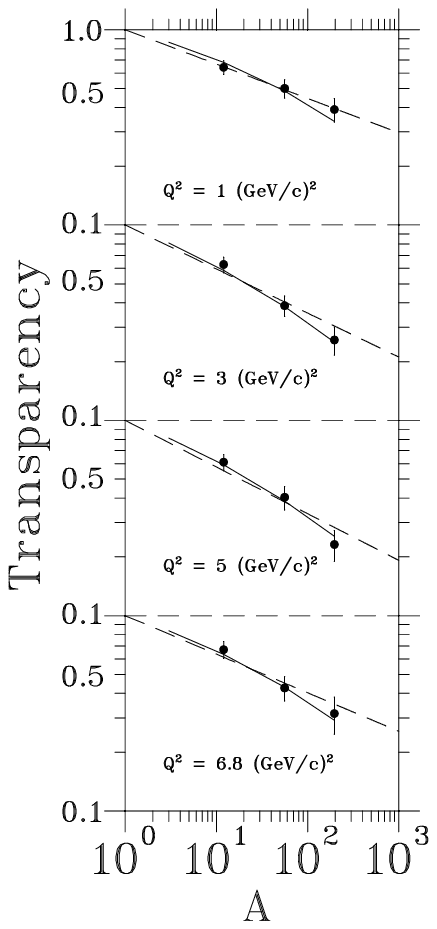}
\vbox{\footnotesize Nuclear transparency (with total errors)
as a function of $A$ for each $Q^2$. The solid line is a fit using
the
classical attenuation model discussed in the text and the dashed line
is a fit
to $T = A^{\alpha}$.}
\vspace{1cm}

In summary, we have measured the nuclear transparency of the
quasielastic $(e,e'p)$ reaction as a function of the nuclear mass $A$
in the $Q^2$ range of 1--7~(GeV/c)$^2$. The measured transparencies
for all the $A > 1$ targets, as well as $\sigma_{\rm eff}$, are
independent of $Q^2$ for $Q^2>3~\hbox{(GeV/c)}^2$ indicating that we
have seen no evidence of effects associated with Color Transparency.

This work was supported in part by the National Science Foundation,
Grants No. PHY-9014406 and PHY-9114958 (American), PHY-9115574
(Caltech), PHY-9101404 (CSLA), PHY-9208119 (RPI), and PHY-9019983
(Wisconsin), and by the Department of Energy, Contracts No.
W-31-109-ENG-38 (Argonne), DE-FG02-86ER40269 (Colorado),
W-7405-Eng-48 (LLNL), DE-AC02-76ER03069 (MIT), DE-AC03-76SF00515
(SLAC), and DE-FG03-88ER40439 (Stanford). RGM acknowledges the
support of a Presidential Young Investigator Award from NSF. BWF
acknowledges the support of a Sloan Foundation Fellowship.

\clearpage
\begin{table}
\squeezetable
\label{tbl1}
\caption{Kinematics of the experiment.
$E$ is the beam energy, $E'$ and $\theta_e$ are the
momentum and angle setting of the electron spectrometer,
and $\theta_p$ is the angle setting
of the proton spectrometer. The momentum of the proton spectrometer
was set
equal to the virtual photon 3-momentum $q$.
The ${}^1$H data were taken at elastic scattering kinematics with the
same
$E$ and $E'$ as the solid targets.}
\begin{tabular}{dcdddc}
$Q^2$ & Targets & $E$ & $E'$ & $\theta_e$ & $\theta_p$ \\
(GeV/c)$^2$ & & (GeV) & (GeV) & (deg) & (deg) \\
\hline
1.04 & C, Fe, Au & 2.015 & 1.39 & 35.5 & 43.4, 46.2, 49.0, 51.8, \\
& & & & & 54.6\\
1.21 & ${}^2$H & & 1.36 & 38.8 & 35.9, 39.1, 41.3, 43.5, \\
& & & & & 46.7\\
3.06 & C, Fe & 3.188 & 1.47 & 47.7 & 27.7, 30.5, 33.3 \\
& Au & & & & 27.7, 30.5 \\
& ${}^2$H & & & & 27.7 \\
5.00 & C, Fe & 4.212 & 1.47 & 53.4 & 20.9, 22.6\\
& Au & & & & 20.9 \\
& ${}^2$H & & & & 19.5 \\
6.77 & C, Fe & 5.120 & 1.47 & 56.6 & 15.9, 16.7, 17.3 \\
& Au & & & & 16.7 \\
& ${}^2$H & & & & 15.9 \\
\end{tabular}
\end{table}

\widetext
\begin{table}
\label{tbl2}
\caption{Measured transparencies (with total errors)
for C, Fe, and Au. Also shown are the
results of the fits to the $A$-dependence shown in Fig.~2.
$\sigma_{\rm free}$ is the
average of the free p-p and p-n total cross sections
from Ref.~\protect\cite{landolt88}.}
\begin{tabular}{ccccccc}
$Q^2$ & $T_{\rm C}$ & $T_{\rm Fe}$ & $T_{\rm Au}$ & $\alpha$ &
$\sigma_{\rm eff}$ & $\sigma_{\rm free}$\\
(GeV/c)$^2$ & & & & & (mb) & (mb)\\
\hline
1.04&0.64$\pm$0.05&0.50$\pm$0.05&0.39$\pm$0.05&-0.18$\pm$0.02&22$\pm$
3&37$\pm$4 \\
3.06&0.63$\pm$0.06&0.39$\pm$0.05&0.26$\pm$0.04&-0.23$\pm$0.02&32$\pm$
3&44$\pm$3 \\
5.00&0.61$\pm$0.06&0.40$\pm$0.06&0.23$\pm$0.04&-0.24$\pm$0.02&32$\pm$
4&43$\pm$3 \\
6.77&0.67$\pm$0.07&0.43$\pm$0.06&0.32$\pm$0.07&-0.20$\pm$0.02&27$\pm$
4&42$\pm$3 \\
\end{tabular}
\end{table}


\begin{references}
\bibitem[*] {atanl} Present address: Argonne National Laboratory,
Argonne, Illinois 60439
\bibitem[\dag] {atpenn} Present address: University of
Pennsylvania, Philadelphia, Pennsylvania 19104
\bibitem[\ddag] {atmaryland} Present address: University of Maryland,
College Park, Maryland 20742
\bibitem[\S] {atboulder} Present address: University of Colorado,
Boulder, Colorado 80309
\bibitem[**] {atmich} Present address: University of Michigan, Ann
Arbor, Michigan 48109
\bibitem[\dag\dag] {atcebaf} Present address: CEBAF, Newport News,
Virginia 23606
\bibitem[\parallel] {atvirgin} Present address: Virginia Union
University,
Richmond, Virginia 23220
\bibitem[\P\P] {atodu} Present address: Old Dominion University,
Norfolk, Virginia 23529
\bibitem[\P] {atcaltech} Present address: California Institute of
Technology, Pasadena, California 91125
\bibitem[\ddag\ddag] {atslac} Present address: SLAC, Stanford,
California 94309
\bibitem[\S\S] {atkent} Present address: Kent State University, Kent,
Ohio 44242
\bibitem{mb82}A. H. Mueller, in
{\it Proceedings of the XVII Rencontre de Moriond, 1982},
edited by J. Tran Thanh Van
(Editions Fronti\`{e}res, Gif-sur-Yvette, France, 1982), p. 13;
S. J. Brodsky, in
{\it Proceedings of the Thirteenth International Symposium on
Multiparticle Dynamics}, edited by W. Kittel, W. Metzger, and A.
Stergiou
(World Scientific, Singapore, 1982), p. 963.
\bibitem{farrar88}G. R. Farrar {\it et al.}, Phys. Rev. Lett. {\bf
61},
686 (1988).
\bibitem{jenmil91}B. K. Jennings and G. A. Miller, Phys. Rev. D {\bf
44},
692 (1991).
\bibitem{benhar92}O. Benhar {\it et al.}, Phys. Rev. Lett. {\bf 69},
881 (1992).
\bibitem{fsz93}L. L. Frankfurt, M. I. Strikman, and M. B. Zhalov,
Penn.
State Univ. preprint, 1993.
\bibitem{ralston88}J. P. Ralston and B. Pire, Phys. Rev. Lett. {\bf
61},
1823 (1988).
\bibitem{carrol88}A. S. Carroll {\it et al.}, Phys. Rev. Lett. {\bf
61},
1698 (1988).
\bibitem{npas84}NPAS Users Guide, SLAC Report No. 269, 1984
(unpublished).
\bibitem{makins94}N. C. R. Makins {\it et al.}, Phys. Rev. Lett. {\bf
72}, 1986
(1994).
\bibitem{fmoug83}S. Frullani and J. Mougey, Advances in Nucl. Phys.
{\bf 14},
1 (1984).
\bibitem{def83}T. De Forest, Nucl. Phys. {\bf A392}, 232 (1983).
\bibitem{gari85} M. F. Gari and W. Kr\"{u}mplemann, Z. Phys. {\bf
A322},
689 (1985).
\bibitem{bonnxx}R. Machleidt {\it et al.}, Phys. Rep. {\bf 149}, 1
(1987).
\bibitem{Quint}E. N. M. Quint, Ph.D. Thesis, U. Amsterdam, 1988.
\bibitem{Negele}J. W. Negele, Phys. Rev. C {\bf 1}, 1260 (1970);
J. W. Negele and D. Vautherin, Phys. Rev. C {\bf 5}, 1472 (1972).
\bibitem{ton94}T. G. O'Neill, Ph.D. thesis, Caltech, 1994.
\bibitem{sick}I. Sick, private communication.
\bibitem{vanorden}J. W. van Orden, W. Truex, and M. K. Banerjee,
Phys. Rev.
C {\bf 21}, 2628 (1980).
\bibitem{ji}X. Ji, private communication.
\bibitem{liuti94}S. Liuti, private communication.
\bibitem{pm}$p_m > 0$ corresponds to the angle of ${\bf p'}$ with
respect to the beam in the horizontal plane being
greater than the angle of ${\bf q}$.
\bibitem{garino92} D. F. Geesaman {\it et al.}, Phys. Rev. Lett. {\bf
63}, 734
(1989); G. Garino {\it et al.}, Phys. Rev. C {\bf 45}, 780 (1992).
\bibitem{seki93}A. Kohama, K. Yazaki, and R. Seki, Nucl. Phys. {\bf
A536},
716 (1992).
\bibitem{rinat93}A. S. Rinat and B. K. Jennings, TRIUMF Preprint,
1993.
\bibitem{nik93} N. N. Nikolaev {\it et al.} Nucl. Phys. {\bf A567},
781 (1994).
\bibitem{deVries} H. de Vries {\it et al.}, At. Data Nucl. Data
Tables
{\bf 36}, 495 (1987).
\bibitem{ralston93} P. Jain and J. P. Ralston, Phys. Rev. D
{\bf 48}, 1104 (1993).
\bibitem{vj92} V. R. Pandharipande and S. C. Pieper, Phys. Rev. C
{\bf 45},
791 (1992).
\bibitem{landolt88} A. Baldini {\it et al.},
{\it Total Cross Sections for Reactions
of High Energy Particles}, Landolt-B\"ornstein, New Series,
Vol. {\bf I/12b},
edited by H. Schopper (Springer-Verlag, 1987).
\end{references}
\end{document}